\documentclass[useAMS,usenatbib,twocolumn]{mn2e}

\usepackage{graphicx,amsmath,amssymb}

\newcommand\apj{{ApJ}}%
\newcommand\apjl{{ApJ}}%
\newcommand\apjs{{ApJS}}%
\newcommand\aap{{A\&A}}%
 \newcommand\actaa{{Acta Astronomica}}%
\newcommand\mnras{{MNRAS}}%
\newcommand\pasp{{PASP}}%
\newcommand\nat{{Nature}}%
\newcommand\uoTable{0.04686(17)} 
\newcommand\sTable{1.822(10)} 
\newcommand\pangleTable{0.3662(36)} 
\newcommand\qTable{0.0284(10)} 
\newcommand\tzeroTable{2455774.2729(54)} 
\newcommand\tETable{23.17(80)} 
\newcommand\prhoTable{$<0.007$} 
\newcommand\fsfbTable{$0.0573(29)$} 
\newcommand\IsTable{$19.83(5)$} 
\newcommand\IbTable{$16.69(2)$} 

\newcommand\q{0.028$\pm$0.001} 
\newcommand\tE{23.2$\pm$0.8} 
\newcommand\prho{<0.007} 

\newcommand\mL{$0.39^{+0.45}_{-0.19}$ $M_{\odot}$} 
\newcommand\dL{$7.56\pm0.91$ ${\rm kpc}$} 
\newcommand\rE{$2.38^{+0.81}_{-0.65}$ AU} 
\newcommand\mP{$11.6^{+13.4}_{-5.6}$ $M_{\rm J}$} 
\newcommand\aP{$4.3^{+1.5}_{-1.2}$ AU} 
\newcommand\SL{$R_{\rm SL}=1.1^{+1.2}_{-0.6}$ AU} 

\title[A ``second generation survey'' microlensing planet]
{\vspace*{-0.9 truecm} MOA-2011-BLG-322Lb: a ``second generation survey'' microlensing planet \vspace*{-0.6 truecm}}
\author[Shvartzvald et al.]
{Y.~Shvartzvald$^1$\thanks{E-mail: yossi@wise.tau.ac.il},
D.~Maoz$^1$, S.~Kaspi$^1$, T.~Sumi$^{2,16}$, A. Udalski$^{3,17}$, A.~Gould$^4$, \and
D.P.~Bennett$^{5,16}$, C.~Han$^{18}$
\newauthor and
\newauthor
F.~Abe$^6$, I.A.~Bond$^7$, C.S.~Botzler$^8$, M.~Freeman$^8$, A.~Fukui$^9$, D.~Fukunaga$^6$, Y.~Itow$^6$,\and
N.~Koshimoto$^2$, C.H.~Ling$^7$, K.~Masuda$^6$, Y.~Matsubara$^6$, Y.~Muraki$^6$, S.~Namba$^2$, \and
K.~Ohnishi$^{10}$, N.J.~Rattenbury$^{8}$, To.~Saito$^{11}$, D.J.~Sullivan$^{12}$, W.L.~Sweatman$^7$,\and
D.~Suzuki$^2$, P.J.~Tristram$^{13}$, K.~Wada$^2$, P.C.M.~Yock$^{8}$,\and
(MOA~collaboration),
\newauthor and
\newauthor
J. Skowron$^3$, S. Koz{\l}owski$^3$, M.K. Szyma{\'n}ski$^3$, M. Kubiak$^3$, G. Pietrzy{\'n}ski$^{3,14}$,\and
I. Soszy{\'n}ski$^3$, K. Ulaczyk$^3$, {\L}. Wyrzykowski$^{3,15}$, R. Poleski$^{3,4}$, P. Pietrukowicz$^3$ \and
(OGLE~collaboration)\\
\\
$^1$ School of Physics and Astronomy, Tel-Aviv University, Tel-Aviv 69978, Israel\\
$^2$ Department of Earth and Space Science, Osaka University, Osaka 560-0043, Japan\\
$^3$ Warsaw University Observatory, Al.~Ujazdowskie~4, 00-478~Warszawa, Poland\\
$^4$ Department of Astronomy, Ohio State University, 140 West 18th Avenue, Columbus, OH 43210, United States of America\\
$^5$ University of Notre Dame, Department of Physics, 225 Nieuwland Science Hall, Notre Dame, IN 46556-5670, USA\\
$^6$ Solar-Terrestrial Environment Laboratory, Nagoya University, Nagoya, 464-8601, Japan\\
$^7$ Institute of Information and Mathematical Sciences, Massey University, Private Bag 102-904, North Shore Mail Centre, Auckland, New Zealand\\
$^8$ Department of Physics, University of Auckland, Private Bag 92-019, Auckland 1001, New Zealand\\
$^9$ Okayama Astrophysical Observatory, National Astronomical Observatory of Japan, Asakuchi, Okayama 719-0232, Japan\\
$^{10}$ Nagano National College of Technology, Nagano 381-8550, Japan\\
$^{11}$ Tokyo Metropolitan College of Aeronautics, Tokyo 116-8523, Japan\\
$^{12}$ School of Chemical and Physical Sciences, Victoria University, Wellington, New Zealand\\
$^{13}$ Mt. John University Observatory, P.O. Box 56, Lake Tekapo 8770, New Zealand\\
$^{14}$ Universidad de Concepci{\'o}n, Departamento de Astronomia, Casilla 160--C, Concepci{\'o}n, Chile\\
$^{15}$ Institute of Astronomy, University of Cambridge, Madingley Road, Cambridge CB3 0HA, UK\\
$^{16}$ Microlensing Observations in Astrophysics (MOA) Collaboration\\
$^{17}$ Optical Gravitational Lens Experiment (OGLE) Collaboration\\
$^{18}$Department of Physics, Chungbuk National University, Cheongju 371-763, Republic of Korea
\vspace*{-0.5 truecm}
}
\date{Submitted 2013 September 30 \vspace*{-0.5 truecm}}

\begin{document}
\maketitle


\begin{abstract}
 \noindent
Global "second-generation" microlensing surveys aim to discover and characterize extrasolar planets and their frequency,
by means of round-the-clock high-cadence monitoring of a large area of the Galactic bulge, in a controlled experiment.
We report the discovery of a giant planet in microlensing event MOA-2011-BLG-322.
This moderate-magnification event, which displays a clear anomaly induced by a second lensing mass, was inside the footprint of our 
second-generation microlensing survey, involving MOA, OGLE and the Wise Observatory. The event was observed by the survey groups,
without prompting alerts that could have led to dedicated follow-up observations.
Fitting a microlensing model to the data, we find that the timescale of the event was
$t_{\rm E}=$\tE~{\rm days}, and the mass ratio between the lens star and its companion is $q=$\q.
Finite-source effects are marginally detected, and upper limits on them help break some of the degeneracy in the system parameters.
Using a Bayesian analysis that incorporates a Galactic structure model, we estimate the mass of the lens
at \mL, at a distance of \dL.
Thus, the companion is likely a planet of mass \mP,
at a projected separation of \aP, rather far beyond the snow line.
This is the first pure-survey planet reported from a second-generation microlensing survey,
and shows that survey data alone can be sufficient to characterize a planetary model.
With the detection of additional survey-only planets, we will be able to constrain the frequency
of extrasolar planets near their systems' snow lines.

\end{abstract}


\section{INTRODUCTION}
\label{sec:intro}

The discovery of thousands of extrasolar planets ranks among the most exciting scientific developments of the past decade.
The majority of those exoplanets were detected and characterized by the transit and radial velocity methods, which favor the 
detection of massive planets in close orbits around their hosts, stars at distances within a few hundred~parsec.
Microlensing, in contrast, can reveal planets down to Earth mass and less (\citealt{Bennett.1996.A}),
at larger orbits - about 1 to 10 AU, which is where the ``snowline'' is located, and beyond which giant planets
are expected to form according to planet formation models (\citealt{Ida.2005.A}).
Microlensing enables the detection of planets around all types of stars at distances as far as the Galactic center,
and even planets unbound from any host star (\citealt{Sumi.2011.A}).
Although planets discovered with microlensing still number in the few tens,
several teams (e.g. \citealt{Gould.2010.B,Sumi.2010.A,Cassan.2012.A})
have attempted to estimate the frequency of planets at these separations. 
\citet{Gould.2010.B} estimated from high-magnification events a $\sim1/6$ frequency of Solar-like systems.
\citet{Sumi.2010.A} found that Neptune-mass planets are three times more common than Jupiters beyond the snowline.
\cite{Cassan.2012.A}, concluded that, on average, every star in the Galaxy hosts a snowline-region planet.
Moreover, the fact that two out of the 20 planetary systems discovered by microlensing,
OGLE-2006-BLG-109Lb,c (\citealt{Gaudi.2008.A}) and OGLE-2012-BLG-0026Lb,c (\citealt{Han.2013.A}), host 2 planets, suggests that multiple systems are common,
as also indicated at smaller separations by transit data from {\it Kepler} (\citealt{Howard.2013.A}).
Over the past decade, microlensing planet discoveries have largely come from observing campaigns in which specific,
high-magnification ($A\gtrsim 100$), events are followed intensively by networks of small telescopes, in order to detect
and characterize planetary anomalies in the light curves. 
High magnification events are very sensitive to planets near their snowlines, but they are rare events. Furthermore,
the inhomogeneous social process through which potentially high magnification events are alerted and followed up complicates their use
for statistical inferences on planet frequency. 

Microlensing surveys have been in transition into the so-called ``second generation'' phase (\citealt{Gaudi.2009.A}),
wherein a large area of the Galactic bulge is monitored continuously,
with cadences high enough to detect planetary anomalies without any follow-up observations or changes to the observing sequence
(e.g. switching to a higher cadence).
The first such survey began in 2011 and combines three groups: OGLE (Optical Gravitational Lensing Experiment) -- observing from Chile (\citealt{Udalski.2009.A}),
MOA (Microlensing Observations in Astrophysics) -- observing from New-Zealand (\citealt{Sumi.2003.A}), and the Wise survey observing from Israel (\citealt{Shvartzvald.2012.A}).
\cite{Shvartzvald.2012.A} have simulated the results that can be expected from this "controlled experiment", and its potential to measure
the abundance of planetary systems. They found that the overall planet detection efficiency for the survey is $\sim20\%$.
In the 2011 season there was a total of 80 events inside the high-cadence survey footprint that is common to all three groups
(an additional 218 events were observed by only two of the groups). Of those 80 events, at least three showed a clear planetary anomaly:
MOA-2011-BLG-293 (\citealt{Yee.2012.A}),  MOA-2011-BLG-322 (this paper) and OGLE-2011-BLG-0265 (in preparation).
Accounting for our detection efficiency, these results imply $\sim1/5$ frequency of planetary systems,
which is in line with previous estimates by \cite{Gould.2010.B}.

In this paper, we present the analysis of MOA-2011-BLG-322Lb.
This is the first planetary microlensing event that is detected and analyzed based solely on second-generation survey data.
In principle, OGLE-2003-BLG-235/MOA-2003-BLG-53Lb (\citealt{Bond.2004.A}) and MOA-2007-BLG-192Lb (\citealt{Bennett.2008.A})
were also discovered and characterized based only on MOA and OGLE data.
This was fortuitously possible despite the sparse sampling of first generation surveys.
MOA-2011-BLG-293Lb (\citealt{Yee.2012.A}) was also characterizable by survey-only data, but it did include a large amount of
non-survey data that were prompted by alerts, following early realization of the event's high magnification. 
We describe the observations by the three survey groups in Section \ref{sec:obs}. In Section \ref{sec:model} we present
the binary microlensing model fitted to the event. A Bayesian analysis estimating the physical properties of the system
is presented in Section \ref{sec:Physical}, and we discuss our results in Section \ref{sec:Discussion}.

\begin{figure*}
\begin{minipage}{\textwidth}
\center
\label{fig:model}
\begin{tabular}{c}
\includegraphics[width=1\textwidth]{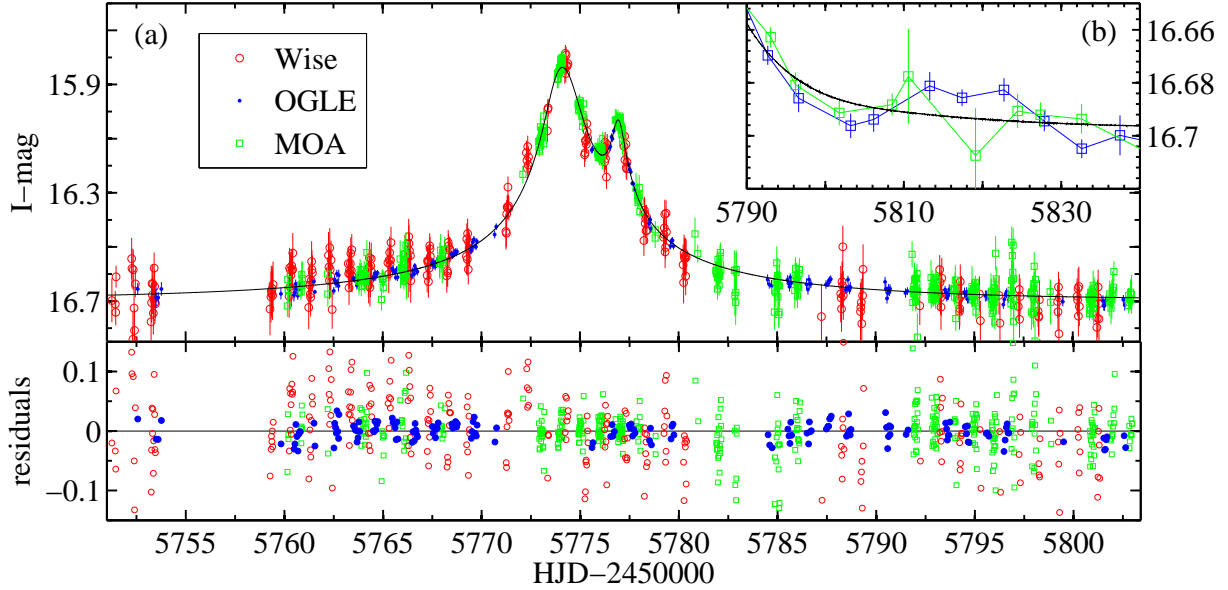}
\end{tabular}
\caption{Inter-calibrated light curve for the 3 datasets: MOA -- green, OGLE -- blue, Wise -- red.
The curve is our best-fit binary lens model (bottom panel shows the residuals from the model).
The apparent magnification is only $\sim 1$ mag due to the large blending fraction. (b) Binned OGLE and MOA data
showing a $\sim0.02$ mag bump, 50 days after the peak of the event, which may be real, although similar features can be found
over the light-curve baseline.}
\end{minipage}
\end{figure*}

\section{Observational data}
\label{sec:obs}

The microlensing event MOA-2011-BLG-322 was first detected on June 30, 2011, 16:15 UT by MOA,
who operate the 1.8m MOA-II telescope at the Mt John Observatory in New Zealand.
The source is located inside the second-generation survey footprint, at RA = 18:04:53.6, Dec = $-$27:13:15.4 (J2000.0).
Thus the event was also observed by the Wise team with the 1m telescope at the Wise Observatory in Israel,
and by the OGLE team with the 1.3m Warsaw telescope at the Las Campanas Observatory in Chile.
It was discovered independently by the OGLE early warning system (EWS, \citealt{Udalski.2003.A}) and designated as OGLE-2011-BLG-1127.
Since the event was not identified as interesting in real time (although MOA noted some anomalous behavior),
the survey teams continued their regular observing cadences for this field.
The observational information for each group (filter, cadence, exposure time) is summarized in Table~1.
The event was not observed by most of the microlensing follow-up groups such as RoboNet-II (\citealt{Tsapras.2009.A}), MiNDSTEp (\citealt{Dominik.2010.A}),
or PLANET (\citealt{Albrow.1998.A}) since it was not very bright at baseline, and had an apparently low magnification ($\sim 1$ mag,
although we show in Section \ref{sec:model}, below, that this is due to blending, and the true magnification was a moderate $A=23$).
We note that Farm Cove Observatory from $\mu$Fun (\citealt{Gould.2008.A}) tried to observe the event for a couple of nights
after the anomaly, but the data were too noisy to be of use.

OGLE and MOA data were reduced by their standard difference image analysis (DIA) procedures (\citealt{Bond.2001.A,Udalski.2003.A}).
The Wise data were reduced using the pySis DIA software (\citealt{Albrow.2009.A}).
The MOA and Wise fluxes were aligned to the OGLE $I$-band magnitude scale, and inter-calibrated to the microlensing model
(see Section \ref{sec:model}).
As further described below, this event includes a large amount of ``blended light'', which could be due to unrelated stars projected
near the source and lens stars, and/or due to the lens star itself. Re-reduction of the pipeline-reduced data from each observatory,
including centroid alignment and correction for a trend of DIA flux with seeing width for MOA and Wise
data (seen for this source in baseline data, before the event),
corrected a number of measurements with systematic errors in the observatories' pipeline reductions. 

Figure~1 shows the observed light curve, with a clear deviation from symmetric, point-mass microlensing (\citealt{Paczynski.A.1986}).
We therefore proceed to the next level of complexity and attempt to model this event as a binary lens.

\begin{table}
\center
\caption{Observational Summary}
\begin{tabular}{l|c|c|c}
\hline
\hline
Group & Filter & Cadence & Exp. time \\
 & & [min]  & [sec] \\
\hline

MOA & MOA-Red$^*$ & 18 & 60 \\

Wise & $I$ & 30 & 180\\

OGLE & $I$ & 60 & 100\\

\hline
\end{tabular}
\\
$^*$equivalent to $R+I$.
\label{table:data}
\end{table}


\section{Microlensing Model}
\label{sec:model}

The basic binary-lens microlensing model requires seven parameters.
Three are the standard ``Paczynski'' parameters: $t_{\rm E}$ -- the event time scale,
i.e. the time it takes the source to cross the Einstein radius of the primary mass; $u_{\rm 0}$ -- the impact parameter,
i.e. the minimum angular separation, in units of the Einstein angle, between the source and lens;
and $t_{\rm 0}$ -- the epoch of minimum separation.
We note that, in our model, these three parameters are with respect to the primary lens star (and not, e.g., with respect to 
the center of mass of the lens and its companion). The Einstein radius is $\theta_E\cdot D_L$, where the Einstein angle is defined through
\begin{equation}
\theta_E^2 = \kappa M \pi_{\rm rel},~\kappa = \frac{4G}{c^2{\rm AU}},
\end{equation}
where $M$ is the mass of the lens star, and the relative parallax is
\begin{equation}
\pi_{\rm rel}={\rm AU}(\frac{1}{D_L}-\frac{1}{D_S}),
\end{equation}
where $D_S$, and $D_L$ are the distances to the source and lens stars, respectively.

\begin{figure}
\label{fig:caustics}
\center
\begin{tabular}{c}
\includegraphics[width=0.5\textwidth]{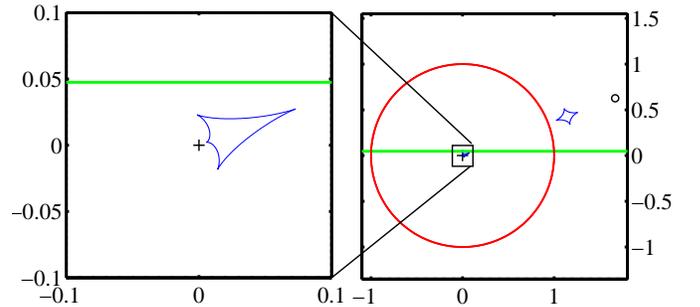}
\end{tabular}
\caption{The trajectory of the source (green line) relative to the ``central'' and ``planetary'' caustics (blue curves).
The host star position is at the center (black cross) and the planet location is indicated by the black circle.
The axes are in units of the Einstein angle, shown as a red circle. Left panel is a zoom in on the central region of the right panel.}
\end{figure}

The possibility of finite source effects due to the non-zero size of the source star is included by allowing a source of angular radius $\rho$,
relative to $\theta_E$, assuming a limb-darkened profile with the ``natural'' coefficient $\Gamma$ (\citealt{Albrow.1999.A}).
Although the source color cannot be measured directly, it is likely a main-sequence G-type star (see below).
We therefore estimate the limb-darkening coefficients from \cite{Claret.2000.A}, using effective temperature $T_{\rm eff}=5750$ K,
and gravity ${\rm log}~g=4.5$, to be $\Gamma_I=0.43$ for OGLE and Wise, and $\Gamma_{R/I}=0.47$ for MOA. 

Three additional parameters are introduced for the companion: $q$ -- the mass ratio between the secondary and the primary; 
and $(s, \alpha)$ -- the two-dimensional projected position of the secondary in the lens plane,
relative to the primary position. These are given by means of a distance, in units of the Einstein radius, and an angle,
measured counter-clockwise from the source trajectory in the lens plane.
In addition to these seven physical parameters, there are two calibration parameters for each dataset $i$, representing the flux
from the source star ($f_{{\rm s,}i}$) and the blended flux $f_{{\rm b,}i}$ from other stars in the line of sight, including the lens.
For a true magnification $A(t)$ at time $t$, the observed flux $F(t)$ is therefore
\begin{equation}
 F(t) = f_{{\rm s,}i}\cdot A(t) + f_{{\rm b,}i}
\end{equation}

In order to solve for the magnification of the binary-lens model, we use the ray-shooting light-curve generator
described in \cite{Shvartzvald.2012.A}.
Briefly, we construct a trial model of the binary lens with a given choice of the parameters in the problem.
We divide the lens plane onto a grid and use the lens equation directly to map the lens plane onto the source plane.
By calculating the entire source trajectory at once, and by using an adaptive grid that increases the lens-plane resolution
around the image positions, we achieve fast computation times.
The magnification is then the ratio of the summed solid angles subtended by all the images in the lens plane to that of the solid angle
subtended by the source.

\begin{table}
\center
\caption{Microlensing model}
\begin{tabular}{l|c}
\hline
\hline

$t_{\rm 0}$ [{\rm HJD}] & \tzeroTable  \\

$u_{\rm 0}$ & \uoTable \\

$t_{\rm E}$ [{\rm days}]& \tETable \\

$s$ & \sTable \\

$\alpha$ [{\rm rad}]&  \pangleTable \\

$q$ &  \qTable \\

$\rho$ & \prhoTable \\

$f_s/f_b$ [OGLE]& \fsfbTable \\

$I_{\rm source}$ [mag] & \IsTable \\

$I_{\rm blend}$ [mag] & \IbTable \\

\hline

$c_{\rm moa}$ & 1.50 \\

$c_{\rm wise}$ & 1.25 \\

$c_{\rm ogle}$ & 1.57 \\

\hline
\end{tabular}
\label{table:model}
\end{table}

A Markov-chain Monte-Carlo (MCMC) search of the parameter space, using Gibbs sampling, is used to find the best-fit solution and its uncertainty.
For every trial model, the calibration parameters were found analytically by means of a linear least squares minimization.
Following previous microlensing event analyses (see e.g. \citealt{Yee.2012.A}), the flux errors, $\sigma'_i$,
were re-normalized such that $\chi^2$ per degree of freedom (DOF) equals unity, using
\begin{equation}
\sigma'_i = c_i \sqrt{\sigma_i^2 + e_{{\rm min,}i}^2},
\end{equation}
where $\sigma_i$ is the pipeline-reported flux errors for each group, and $c$ and $e_{\rm min}$ are the re-normalization coefficients.
$e_{\rm min}$ is a systematic photometric error floor that dominates when the source is very bright and the Poisson errors become small.

Figure~1 shows the inter-calibrated light-curve of the event and our best-fit model.
The best-fit parameters are given in Table~2, along with the re-normalization coefficients, $c_i$, for each group ($e_{\rm min}=0$ for all groups).
The time scale of the event, $t_{\rm E}=$ \tE~{\rm days}, suggests a sub-solar-mass primary lens.
The mass ratio between the primary and the secondary is $q=$\q, about 30 times the Jupiter/Sun ratio,
near the brown-dwarf/planetary border for an M-type host star. 
As mentioned above, almost 95\% of the flux is blended light
due to a star near the line-of-sight. We discuss this result and its implications in Section~\ref{sec:Discussion}.

We note the existence of a small, $\sim 0.02$~mag, ``bump'' in the light curve, at $\sim t_0 + 50$ d, with duration of about 20 days.
Figure~1b shows the 5-day-binned OGLE and MOA data over this period.
The possibly similar behavior in OGLE and MOA data lends some credence to the reality of this feature. On the other hand, its
amplitude is similar to those of several other baseline fluctuations, and thus no definite conclusions can be drawn about it.

The trajectory of the source relative to the caustic structure is shown in Figure~2.
Since the source passes near the central caustic, we have checked for the so-called $s \leftrightarrow s^{-1}$ degeneracy,
that often occurs with anomalies dominated by the central caustic in the case of high-magnification events (\citealt{Griest.1998.A}).
However, the close solution is disfavored by $\bigtriangleup\chi^2=85$, and we adopt the wide solution. 

\begin{figure}
\label{fig:cmd}
\center
\begin{tabular}{c}
\includegraphics[width=0.5\textwidth]{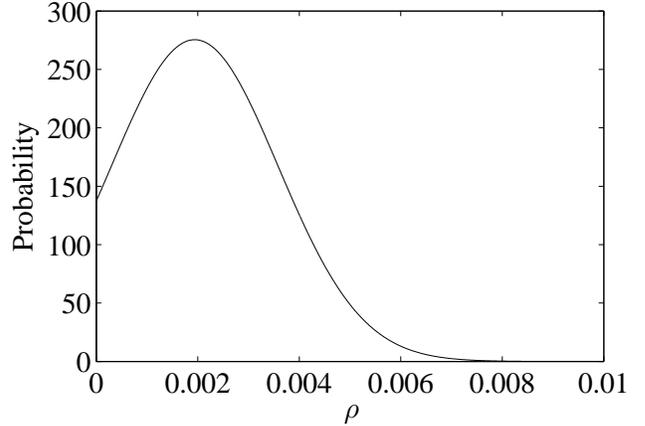}
\end{tabular}
\caption{The recovered probability distribution for the finite source size,
setting an upper limit of $\rho\prho$, and showing that the data are consistent with a point source.
The distribution is used to construct a prior for our Bayesian analysis.}
\end{figure}

In addition, since the event shows no evidence for a caustic crossing,
we have checked if the event could be explained by a single lens and a binary source.
The magnification of each source is defined by the standard ``Paczynski'' parameters, but with a single time scale for both sources, $t_{\rm E}$.
For each dataset we introduce an additional flux coefficient,
\begin{equation}
F(t) = f_{{\rm s1,}i}\cdot A_1(t) + f_{{\rm s2,}i}\cdot A_2(t) + f_{{\rm b,}i}.
\end{equation}
We find that such a model fits the data as well as the binary-lens model (formally slightly better, $\Delta\chi^2=5$,
for 1703 degrees of freedom). 
\cite{Gaudi.1998.B} has noted that a binary-source model and a planetary model can be distinguished by the color difference expected for two sources
of unequal luminosity.
Recently, \cite{Hwang.2013.A} have used this method to resolve this degeneracy for microlensing event MOA-2012-BLG-486, and found that it was a binary-source event.
For our event, the apparent unlensed magnitudes of two sources would be $20.06\pm0.08$ and $21.86\pm0.07$, corresponding to G- and K-type main sequence stars,
respectively, at the distance of the bulge.
For such stars, this would predict a MOA$-$OGLE color difference between the two sources,
$-2.5 {\rm log}_{\rm 10}[(f_{\rm s1,MOA}/f_{\rm s1,OGLE})/(f_{\rm s2,MOA}/f_{\rm s2,OGLE})]$, of $-0.2$.
This is inconsistent with the recovered color difference from the binary source model, of only $-0.03\pm0.04$ mag, and rules out the binary-source model.

In many anomalous microlensing events (e.g. \citealt{Udalski.2005.A, Gaudi.2008.A, Muraki.2011.A, Han.2013.A, Kains.2013.A}),
high-order effects, such as microlens parallax and finite-source effects, can break (or partially break) the degeneracies among the physical parameters
that determine $t_{\rm E}$. However, since the event duration was short and with moderate magnification, the amplitude of the microlens parallax effect is small,
and including it does not improve the fit significantly ($\Delta\chi^2=7$).
Our modeling sets an upper limit on the Einstein-radius-normalized source radius of $\rho\prho$ (3$\sigma$ level).
The best-fit model has $\rho=0.002$, but the data are also consistent with a point source at the $\sim1\sigma$ level ($\Delta\chi^2=1.4$).
The probability distribution for the finite-source size, recovered from the MCMC chain, is shown in Figure~3.

To try to set further constraints on the source and lens properties, we construct a color-magnitude diagram (CMD) of objects
within 90$''$ of the event's position (Figure~4), using OGLE-III (\citealt{Udalski.2008.A}) calibrated  $V$-band and $I$-band magnitudes
of images, taken before the event (there were no $V$-band images during the event). 
We estimate the position of the ``red clump'' at 
$(V-I,I)_{\rm cl}=(1.94,15.45)$ and compare it to the unreddened values derived by \cite{Nataf.2013.A}
for the Galactic coordinates of the event, (l,b)=(3.6,$-$2.8). \cite{Nataf.2013.A} and \cite{Bensby.2011.A} find $(V-I,I)_{\rm cl,0}=(1.06,14.36)$,
i.e. a line-of-sight $I$-band extinction to the red clump of $A_I=1.09$, and a distance modulus of $14.48\pm0.24$ mag,
or $7.9\pm0.9$ kpc.

\begin{figure}
\label{fig:cmd}
\center
\begin{tabular}{c}
\includegraphics[width=0.5\textwidth]{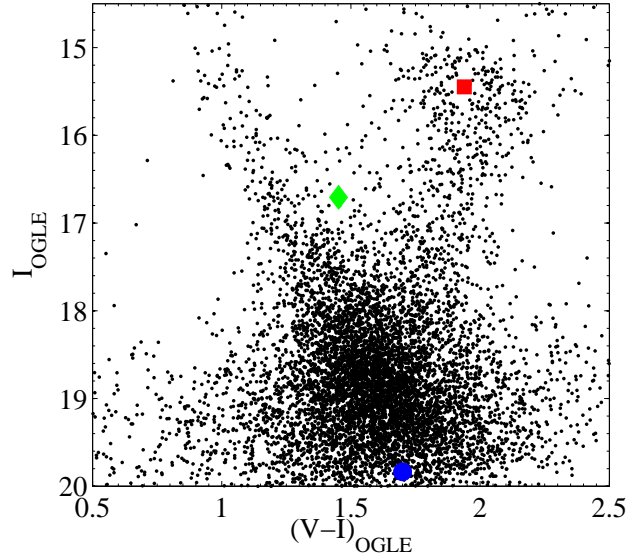}
\end{tabular}
\caption{OGLE-III calibrated CMD of stars within 90$''$ of the event's position.
The red square is an estimate of the red clump centroid.
The green diamond shows the blend star and the blue circle represent the source star - its OGLE $I$-mag is measured
from the microlensing model, and its unknown color is plotted at the location corresponding
to a G-type dwarf at the distance of the Galactic bulge.}
\end{figure}

\begin{figure*}
\begin{minipage}{\textwidth}
\center
\label{fig:Bayesian}
\center
\begin{tabular}{c}
\includegraphics[width=0.3\textwidth]{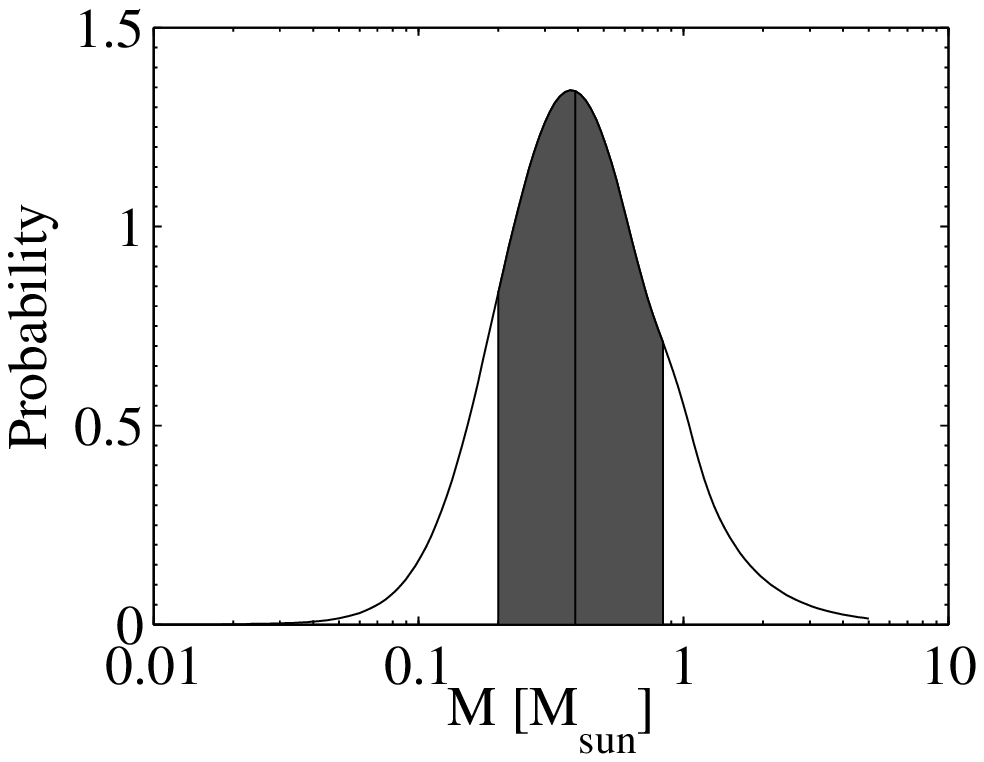}
\includegraphics[width=0.3\textwidth]{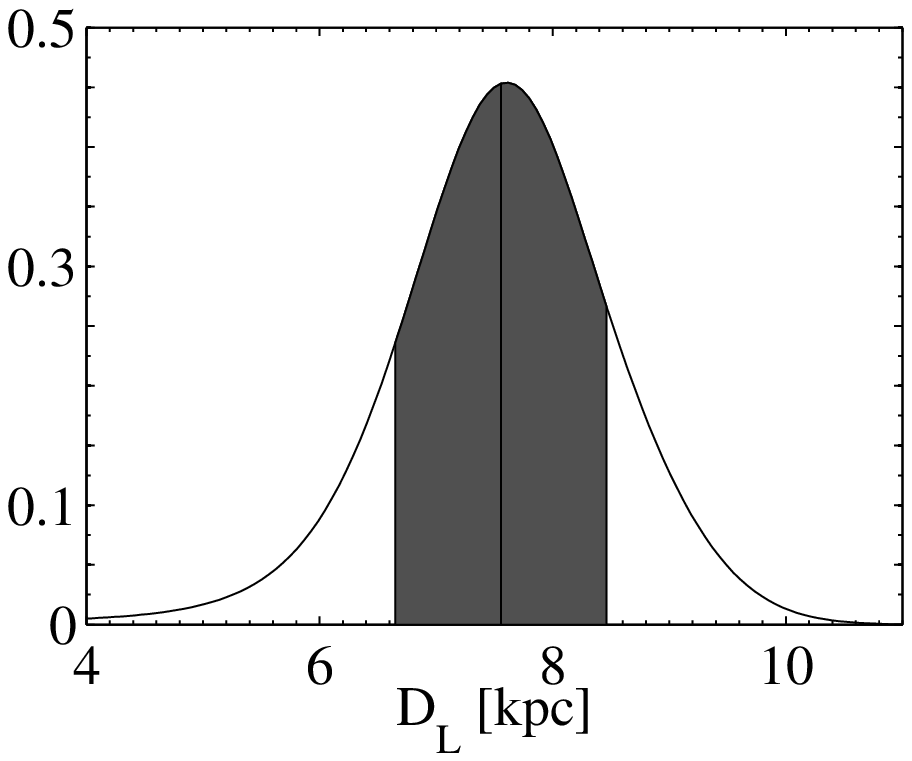}
\includegraphics[width=0.3\textwidth]{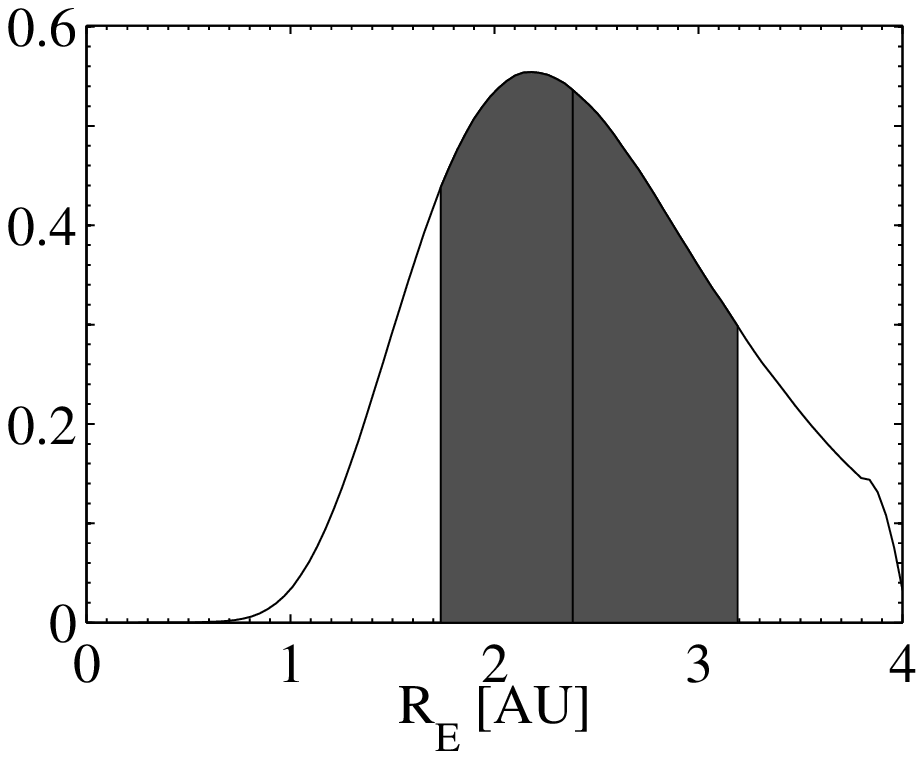}
\end{tabular}
\caption{Bayesian posterior distributions for the physical parameters of the lens.
The vertical lines indicate the median and the shaded areas are the 68\% confidence limits. 
{\it Left}: Lens mass probability density indicating that the lens is probably an M-type star.
{\it Center}: Lens distance probability density suggesting that the system is located in the Galactic bulge.
{\it Right:} Einstein radius probability density - the relatively small size is a consequence of the short $t_{\rm E}$.}
\end{minipage}
\end{figure*}

Due to the strong blending near the lens position, the source star is not detectable in images before the event,
and thus its color is unknown. However, the inferred source $I$-mag from the microlensing model, using the base flux of the blend star and
accounting for the blending fraction, is $19.83\pm 0.05$.
Assuming the source star is at the same distance as the red clump, its absolute magnitude is $M_I = 4.26\pm 0.25$,
i.e. likely a G-type main-sequence star of angular size $\theta_*\approx0.6~\mu$as (\citealt{Dotter.2008.A}).
Combining this with our lens model's upper limit on $\rho$ ($\prho$), a lower limit on the Einstein angle is
\begin{equation}
\theta_E>0.085~{\rm mas}.
\end{equation}
For the best fit value of $\rho=0.002$, the Einstein angle is $\theta_E\approx0.3~{\rm mas}$.
The corresponding proper motion is
$\mu=\theta_E/t_E\approx4.7~{\rm mas/year}$.

In principle, some or all of the bright ``blend light'' in this event could be due to the lens star itself.
Its isolated position on the CMD, at $(V-I,I)_{\rm blend}=(1.45,16.7)$ (see Figure~4), $\sim1$ mag brighter than the general track of
the stars in this region of the diagram, suggests that it might be a foreground disk star that suffers from relatively less extinction.
On the other hand, our Bayesian analysis, below, suggests such a nearby lens is highly unlikely, and therefore
the blend may be due to one or more unrelated stars seen in projection. In addition, based on long-term OGLE data,
the proper motion of the blend star relative to the majority bulge stars in the field is very small, 2$\pm$1 mas/year,
thus also suggesting it is located in the bulge.
If so, it might be a horizontal branch star.
Another possibility is that this blend is not the lens itself, but a companion to the lens.


\section{Physical parameters - Bayesian Analysis}
\label{sec:Physical}

The physical parameters of the lens and its companion are connected to three of the model parameters: $t_{\rm E}$ -- which involves
the lens mass and distance, and $q$ and $s$ which, given the lens properties, give the companion mass and projected distance to
the lens star, respectively.
Given the limited constraints we can set, in this case, using high-order effects and the CMD, 
we use Bayesian analysis to estimate probabilistically the lens distance and mass, and the Einstein radius
of the event, following previous analyses of microlensing events (e.g. \citealt{Batista.2011.A,Yee.2012.A}).
The prior distribution includes a Galactic stellar structure model, which sets the rate equation for lensing events.
As noted, due to the large amount of blended light, likely unrelated to the lens,
we cannot place strong constraints on the lens mass from the observed flux.

The event rate is
\begin{equation}
\begin{split}
\dfrac{d^5\Gamma}{dD_{\rm L} dD_{\rm S} dM d^2\mu}&\\
 =(2R_{\rm E})& V_{\rm rel} n_{\rm L}(x,y,z) n_{\rm S}(x,y,z) D_{\rm S}^2 f(\mu) g(M),
\end{split}
\end{equation}
where $R_{\rm E}$ is the Einstein radius, and $V_{\rm rel}$ is the lens 
transverse 
velocity relative to the source-observer line of sight.
$V_{\rm rel}$ is related to $\mu$, the lens-source relative proper motion, by  $V_{\rm rel}=\mu D_{\rm L}$.
The local densities of lenses and sources are $n_{\rm L}(x,y,z)$ and $n_{\rm S}(x,y,z)$, and $f(\mu)$ is the
lens-source relative proper motion probability distribution. Finally, $g(M)$ is the lens mass function.
We incorporate our constraints on the Einstein angle, $\theta_E$ (Eq. 5),
by using the MCMC probability distribution for $\rho$, to construct the prior distribution for $\theta_E$, which is assigned
to each combination of $M$, $D_L$, and $D_S$ in the rate equation.

We adopt the Galactic model of \cite{Han.1995.A,Han.2003.A}, which reproduces well the observed statistical distribution of properties of 
microlensing events. The stellar density, $n(x,y,z)$, includes a cylindrically symmetric disk,
a bulge and a central bar (for specific model parameters see table 2 in \citealt{Batista.2011.A}).

The proper motion probability distribution, $f(\mu)$, is a two dimensional (Galactic North and East directions) Gaussian distribution,
with expectation value
\begin{scriptsize}\end{scriptsize}
\begin{equation}
 \mu_{\rm exp}=\dfrac{V_{\rm L}-(V_\odot+V_\oplus)}{D_{\rm L}} - \dfrac{V_{\rm S}-(V_\odot+V_\oplus)}{D_{\rm S}}.
\end{equation}
The Sun's velocity, consisting of a random and a circular velocity component, is
$(V_{\odot,{\rm N_{gal}}},V_{\odot,{\rm E_{gal}}})=(7,12)+(0,230)~{\rm km~s^{-1}}$.
The Earth's velocity, as seen by the Sun during the event, was
$(V_{\oplus,{\rm N_{gal}}},V_{\oplus,{\rm E_{gal}}})=(-19.8,9.2)~{\rm km~s^{-1}}$.
$V_{\rm L}$ and $V_{\rm S}$ are the expectation values for the  lens and source velocities, which differ for disk and bulge lenses.
We adopt means and standard deviations for the disk velocities of $(V_{\rm disk,N_{gal}},V_{\rm disk,E_{gal}})=(0,220)~{\rm km~s^{-1}}$,
and $(\sigma_{\rm disk,N_{gal}},\sigma_{\rm disk,E_{gal}})=(20,30)~{\rm km~s^{-1}}$, respectively.
For the bulge, $(V_{\rm bulge,N_{gal}},V_{\rm bulge,E_{gal}})=(0,0)~{\rm km~s^{-1}}$
with $(\sigma_{\rm bulge,N_{gal}},\sigma_{\rm bulge,E_{gal}})=(100,100)~{\rm km~s^{-1}}$.

For the mass function, we follow \cite{Dominik.2006.A} and use different mass functions for the Galactic disk and bulge,
consisting of power laws and a log-normal distributions in $(M/M_{\odot})$ adopted from \cite{Chabrier.2003.A}.

The posterior probability distributions for the lens mass and distance are found by marginalizing over all other parameters,
and for the Einstein radius by summing the probability for the appropriate combination of $D_L$, $D_S$ and $M$, marginalizing over $\mu$.
Figure~5 shows the results of our Bayesian analysis.
The inferred lens mass is $M=0.39^{+0.45}_{-0.19}$ $M_{\odot}$, and thus the companion is a planet with mass of \mP.
The uncertainties are the 68\% probability range about the median of the probability distribution, which we take as the most likely value.
The system's distance is \dL.
We find that the Einstein radius of the lens star is \rE, so the projected separation of the companion,
$r_{\bot}=s\cdot r_{\rm E}$, is \aP, rather far beyond the location of the snowline at
\SL~[assuming a relation $R_{\rm SL}=2.7(M/M_{\odot})~{\rm AU}$].

\section{Discussion}
\label{sec:Discussion}

\vspace{-0.11cm}
We have presented the detection, via microlensing, of a Jovian planet orbiting around a likely M-type star.
There are three additional microlensing events that have detected Jupiters around M stars
(\citealt{Udalski.2005.A,Batista.2011.A,Poleski.2013.A}), and together they constitute $\sim$1/5 of the planetary systems detected 
to date through microlensing. In all of them, the planet is located beyond the snowline, at distances $\lesssim$5~AU from the host star.
However, within the uncertainties, for two out of the four events (\citealt{Poleski.2013.A} and this work),
the primary could also be a K- or G-type star.
If snowline-region massive planets around M stars are indeed common, this may be in conflict with the two leading
planetary formation scenarios.
According to the ``core accretion'' scenario (e.g. \citealt{Ida.2005.A}), Jovian planets form beyond the
snowlines of their parent stars, but massive planets around M-type stars should be rare (\citealt{Laughlin.2004.A}),
since their formation times are longer than the typical disk lifetime.
In the disk instability planet-formation scenario (e.g. \citealt{Boss.2006.A}) massive planets do form around M stars,
but at distances $\gtrsim$7~AU.

Future high resolution imaging could confirm our results by isolating the light from the lens. In addition, a set of such images,
spread over several years after the event, could measure the relative proper motion between the lens and the source star,
and set stronger constraint on the system parameters.
Since this is a general problem for many of the planets detected by microlensing, a dedicated observing program following
all microlensing planets is essential for the interpretation of those planets, and could give better priors for future Bayesian analysis
of such events. 

Unlike most of the microlensing-detected planets to date, the planet presented here was not detected in real time,
but in a post-season analysis, illustrating the essence and elegance of the second-generation survey principle.
The other planetary events that were inside the collaboration footprint in the 2011 season could have also been fully characterized
by the survey data alone.
The final results of this controlled experiment can thus eventually  help us determine the frequency of snowline planets in the Galaxy.

\section*{Acknowledgments}

We thank J.C.~Yee for stimulating discussions on Bayesian analysis
and M. Albrow for kindly providing the pySis software, which was used for the DIA of the Wise survey data.
The anonymous referee is thanked for comments that improved the presentation.
This research was supported by the I-CORE program of the Planning and Budgeting Committee and the Israel Science Foundation, Grant 1829/12.
DM and AG acknowledge support by the US-Israel Binational Science Foundation.
The OGLE project has received funding from the European Research Council under the European Community's Seventh Framework Programme
(FP7/2007-2013) / ERC grant agreement No. 246678 to AU.
TS acknowledges funding from JSPS 23340044 and JSPS 24253004.


\end{document}